\documentclass[aps,prb,epsf,floatfix,twocolumn,footinbib]{revtex4}
\usepackage{times}
\usepackage{graphicx}
\usepackage{float}
\usepackage{latexsym,amsmath,amssymb,bm,euscript}
\usepackage{color}
\usepackage{subfigure}
\usepackage{epstopdf}
\usepackage{hyperref}

\begin{document}

\title{Quantum phase diagram of the spin-$1$ $J_1-J_2$ Heisenberg model on the honeycomb lattice}
\author{Shou-Shu Gong, Wei Zhu, and D. N. Sheng}
\affiliation{Department of Physics and Astronomy, California State University, Northridge, California 91330, USA}

\begin{abstract}
  Strongly correlated systems with geometric frustrations can host the emergent phases of matter with 
  unconventional properties.
  Here, we study the spin $S = 1$ Heisenberg model on the honeycomb lattice with the antiferromagnetic 
  first- ($J_1$) and second-neighbor ($J_2$) interactions ($0.0 \leq J_2/J_1 \leq 0.5$) by means
  of density matrix renormalization group (DMRG).
  In the parameter regime $J_2/J_1 \lesssim 0.27$, the system sustains a N\'{e}el antiferromagnetic phase. 
  At the large $J_2$ side $J_2/J_1 \gtrsim 0.32$, a stripe antiferromagnetic phase is found.
  Between the two magnetic ordered phases $0.27 \lesssim J_2/J_1 \lesssim  0.32$, we  find
  a \textit{non-magnetic} intermediate region with a plaquette valence-bond order.
  Although our calculations are limited within $6$ unit-cell width on cylinder, 
  we present evidence  that this plaquette state could be a strong candidate for this non-magnetic 
  region in the thermodynamic limit.
  We also briefly discuss the nature of the quantum phase transitions in the system.
  We gain further insight of the non-magnetic phases in the spin-$1$ system by comparing its phase
  diagram with the spin-$1/2$ system.
\end{abstract}

\maketitle

\section{Introduction}

Since Anderson proposed the resonating valence bond theory to
explain the high-temperature superconductivity \cite{anderson1973},
the study  of spin liquid (SL) \cite{rokhsar1988, 2010Natur.464..199B,PhysRevLett.86.1881,PhysRevB.64.064422,PhysRevB.66.205104,
PhysRevB.65.224412,PhysRevLett.94.146805, kitaev2006, yao2007, isakov2011}
in frustrated magnetic systems have been
attracting much attentions for almost $40$ years \cite{RevModPhys.78.17, Science.321.1306}.
In recent years, this field has achieved exciting progresses with identifying
realistic examples of SL states.
Among the various SL candidates, the most promising candidate
is the spin-$1/2$ kagome Antiferromagnet. In experimental side, the strong evidences supporting the
gapless SL have been discovered in the spin-$1/2$ kagome antiferromagnet materials Herbertsmithite
\cite{PhysRevLett.98.077204,PhysRevLett.98.107204,PhysRevLett.103.237201,2012Natur.492..406H}
and Kapellasite \cite{colman2008, PhysRevLett.109.037208, bernu2013, kermarrec2014}. In theoretical studies, the spin-$1/2$ kagome Heisenberg
model has been found to sustain a SL ground state although the nature of the SL is still under debate
between the gapped $Z_2$ SL \cite{PhysRevLett.101.117203, 2011Sci...332.1173Y, PhysRevLett.109.067201,
jiang2012} obtained
from density-matrix renormalization group (DMRG) and the gapless $U(1)$ Dirac SL favored in the variational studies of the
Gutzwiller projected wavefunction \cite{ran2007, iqbal2013, iqbal2014}.
Very recently, by introducing the second and third neighbor interactions \cite{messio2012, 2014NatSR...4E6317G, PhysRevLett.112.137202} or the chiral interactions \cite{2014NatCo...5E5137B} in kagome systems, a gapped chiral spin liquid which
breaks time-reversal symmetry is unambiguously established as the $\nu = 1/2$ fractional quantum Hall 
state through fully characterizing the topological properties \cite{2014NatSR...4E6317G, PhysRevLett.112.137202, 2014NatCo...5E5137B}
of the state.

Besides the kagome systems, the spin-$1/2$ $J_1$-$J_2$ Heisenberg models on the
square and honeycomb lattices have also being considered as the promising
candidates of SL. In particular, the SL on square lattice is considered significant to
understand the high-temperature superconductivity in copper oxide \cite{anderson1973}.
Recently, the long-debated non-magnetic regions in these two models have been studied intensively
as the possible realizations of gapped $Z_2$ 
SL \cite{clark2011, PhysRevB.84.024420, PhysRevB.82.024419, PhysRevB.86.024424}
or gapless SL \cite{hu2013, wang2013}. By performing DMRG calculations on cylinder systems\cite{PhysRevLett.110.127203, PhysRevLett.110.127205, PhysRevB.88.165138, gong2014}, 
it is found that the $Z_2$ SL behaviors in the intermediate region of the both models
appear not stable on the wide systems. Instead, a plaquette valence-bond (PVB) state may 
dominate the non-magnetic regions.

\begin{figure}
\includegraphics[width = 1.0\linewidth]{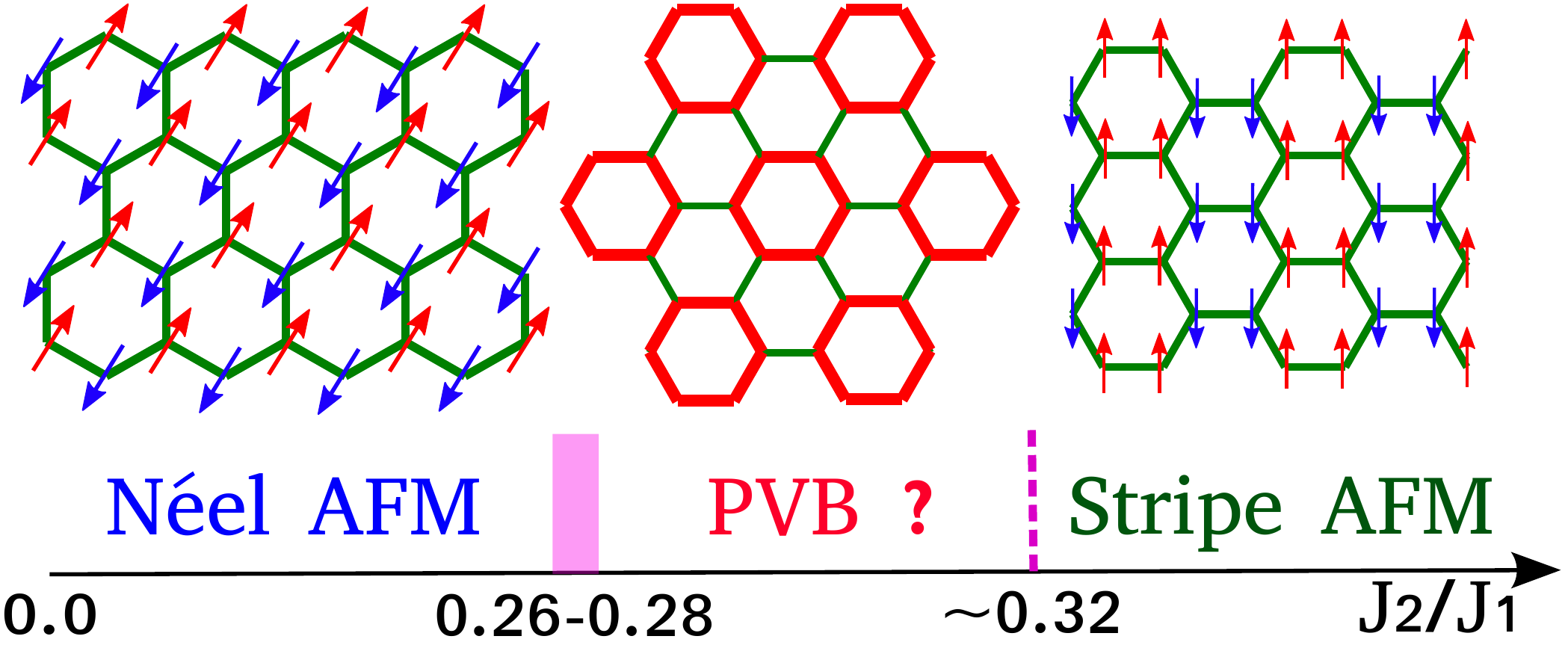}
\caption{Quantum phase diagram of the spin-$1$ $J_1$-$J_2$ Heisenberg model on the honeycomb lattice.
	With increasing $J_2$ coupling, the system has a N\'{e}el AFM phase for $J_2 < 0.26$,
	a stripe AFM phase for $J_2 \gtrsim 0.32$, and an intermediate non-magnetic phase with a
	plaquette valence-bond order in our DMRG calculations. As the finite-size effects on different
	cylinder geometries, the first transition point is estimated at $0.26\sim 0.28$.
} \label{phasediagram}
\end{figure}

The frustrated spin-$1$ magnetic systems on the square and honeycomb lattices are also particularly interesting 
as they may be the parent magnetic systems for  the iron-based superconductivity.
The square Heisenberg models with frustrating further-neighbor couplings and quadratic interactions
\cite{si2008, fang2008, xu2008, jiang2009, yu2012, yu2015} have been studied intensively
using the mean-field analysis to investigate the possible nematic order.
By using DMRG calculations on the spin-$1$ $J_1$-$J_2$ square Heisenberg model \cite{jiang2009},
a non-magnetic phase between the N\'{e}el and the stripe antiferromagnetic (AFM) phase is obtained.
However, the nature of this non-magnetic phase is far from clear although a nematic paramagnetic 
state has been proposed based on the field-theory description \cite{wang2015}. On the other hand,
the frustrated spin-$1$ honeycomb Heisenberg models have not been studies systematically, which may be 
relevant to the spin model for the honeycomb iron-based superconductivity material SrPtAs 
\cite{yoshihiro2011, youn2012, goryo2012, biswas2013, fuscher2014}.
For the spin-$1/2$ $J_1$-$J_2$ honeycomb model
\cite{fouet2001, mosadeq2011, albuquerque2011, mulder2010, cabra2011, zhang2013,
clark2011, xu2011, reuther2011, oitmaa2011, fabio2012, bishop2012, rosales2013,
PhysRevLett.110.127203, PhysRevLett.110.127205, PhysRevB.88.165138, flint2013, ciolo2014},
a N\'{e}el AFM phase and a staggered dimer phase are found at $J_2 \lesssim 0.22$ and $J_2 \gtrsim 0.35$, respectively. 
Between these two phases for $0.25 \lesssim J_2 \lesssim 0.35$,
a PVB phase is identified in DMRG calculations \cite{PhysRevLett.110.127203,
PhysRevLett.110.127205, PhysRevB.88.165138} as the PVB correlation length keeps growing
fast with system width on cylinder \cite{PhysRevLett.110.127205,PhysRevB.88.165138}.
For spin $S=1$ $J_1$-$J_2$ model, the studies are rare and it 
is unclear whether the quantum phases such as the PVB and the staggered dimer
phases would persist with spin magnitude increasing from $1/2$ to $1$, and what kinds of classical
states might emerge in the honeycomb system \cite{PhysRevB.83.184401}.

In this article, we study the spin-$1$ Heisenberg model on the honeycomb lattice
with the frustrating $J_1$-$J_2$ AFM interactions by using the DMRG with
spin rotational $SU(2)$ symmetry \cite{white1992, mcculloch2002}. The Hamiltonian of the model is given as
\begin{equation}
H=J_{1}\sum_{\langle i,j\rangle}\textbf{S}_{i}\cdot\textbf{S}_{j}
+J_{2}\sum_{\langle\langle i,j\rangle\rangle}\textbf{S}_{i}\cdot\textbf{S}_{j},
\end{equation}
where $J_1$ and $J_2$ are the first- and second-neighbor AFM interactions.
We set $J_1$ as the energy scale, and lattice spacing between the nearest-neighbor
sites as the length scale.
Through our $SU(2)$ DMRG calculations
on cylinder systems, we establish a quantum phase diagram as shown in Fig.~\ref{phasediagram}.
By studying the spin correlation function, we find a N\'{e}el AFM phase 
for $0\leq J_2 \lesssim 0.27$. For $0.5 \geq J_2 \gtrsim 0.32$, we find the magnetic ordered
state rather than the staggered dimer state. In this region, the obtained magnetic
order state depends on the cylinder geometry in our finite-size calculations.
By comparing the ground-state bulk energy on different cylinder geometries, we find that the 
stripe AFM state always possesses the lower energy and thus appears to be the ground state in the thermodynamic limit.
Between the two magnetic order phases $0.27 \lesssim J_2 \lesssim 0.32$, the system has a narrow non-magnetic
region with the non-uniform bond energy on wide cylinders. During the increase of the kept states
from $2000$ to $8000$ $SU(2)$ states, a PVB dimer order is found stabilized on the
studied cylinder systems, which suggests that the PVB state is a strong candidate for this
intermediate phase region. Finally, we discuss the nature of the quantum phase transitions in the system
with the help of the bipartite entanglement entropy.

\begin{figure}
  \includegraphics[width = 1.0\linewidth]{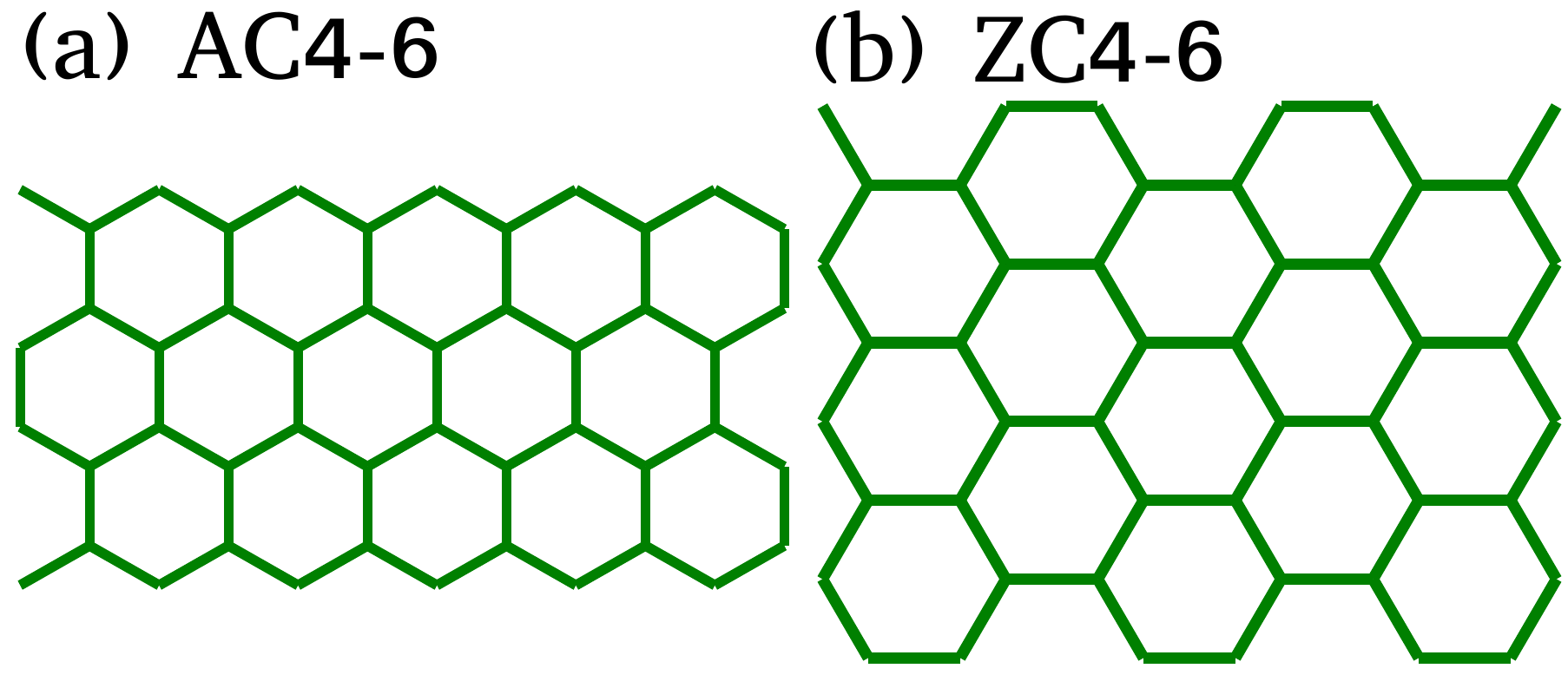}
  \caption{Cylinder geometries used in the DMRG calculations. (a) is an AC4-6 cylinder and (b) is 
	  a ZC4-6 cylinder.} \label{lattice}
\end{figure}

In our DMRG calculations, we study the cylinder systems with width up to $8$ ($6$) unit cells in
the magnetic ordered phases (intermediate phase), by keeping up to $8000$ 
$SU(2)$ states to ensure the convergence. The truncation error is 
controlled below $10^{-6}$ for $L_y = 4$ ($L_y$ is the number of unit cell in the $y$ direction)
cylinder and below $10^{-5}$ for the other calculations. The cylinder geometries are shown in Fig.~\ref{lattice}.
The first cylinder AC$m$-$n$ has the armchair open edges,
where $m$ is the number of two-site 
unit cells along the $y$ direction and $n$ is the number of 
columns along the $x$ direction.
The second cylinder has the zigzag open edges and is denoted as 
ZC$m$-$n$ cylinder.

\section{N\'{e}el AFM phase}

First of all, we study the N\'{e}el AFM phase in the small $J_2$ side.
Due to  the limit of system width in the DMRG calculations for the spin-$1$
system, we do not have enough data of magnetization on different system widths
to estimate the result in thermodynamic limit through extrapolation. Instead,
on the finite-width cylinders, we calculate the spin correlation functions
along the cylinder axis direction (the $x$ direction) and study their decay
behaviors with increasing $J_2$.

\begin{figure}
  \includegraphics[width = 1.0\linewidth]{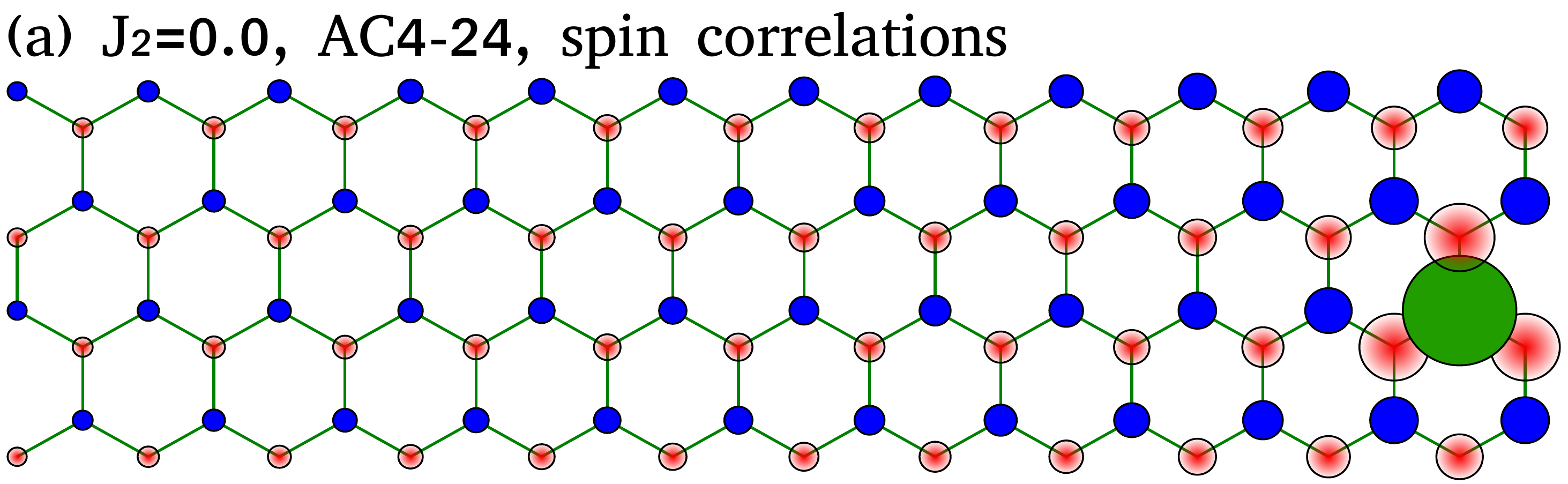}
  \includegraphics[width = 1.0\linewidth]{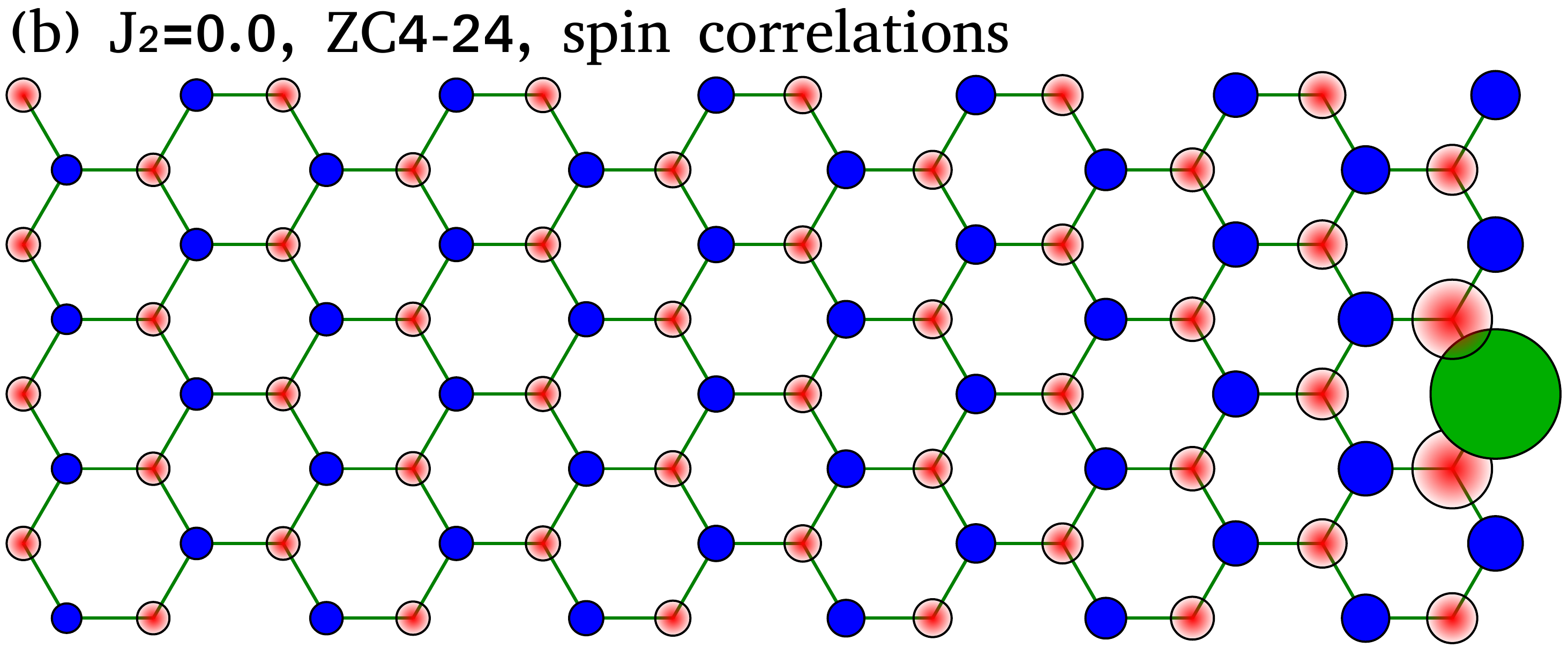}
  \caption{Spin correlation functions in real space for $J_2 = 0.0$ on
	  (a) AC4-24 cylinder and (b) ZC4-24 cylinder. The green site is the
	  reference site in the middle of cylinder, and the blue solid and red shaded circles
	  denote the positive and negative correlations, respectively. 
	  The radius of circle is proportional to the magnitude of correlations.
	  Here, we only show the left half $2\times 4\times 12$ sites.} \label{spinpattern_Neel}
\end{figure}

\begin{figure}
  \includegraphics[width = 1.0\linewidth]{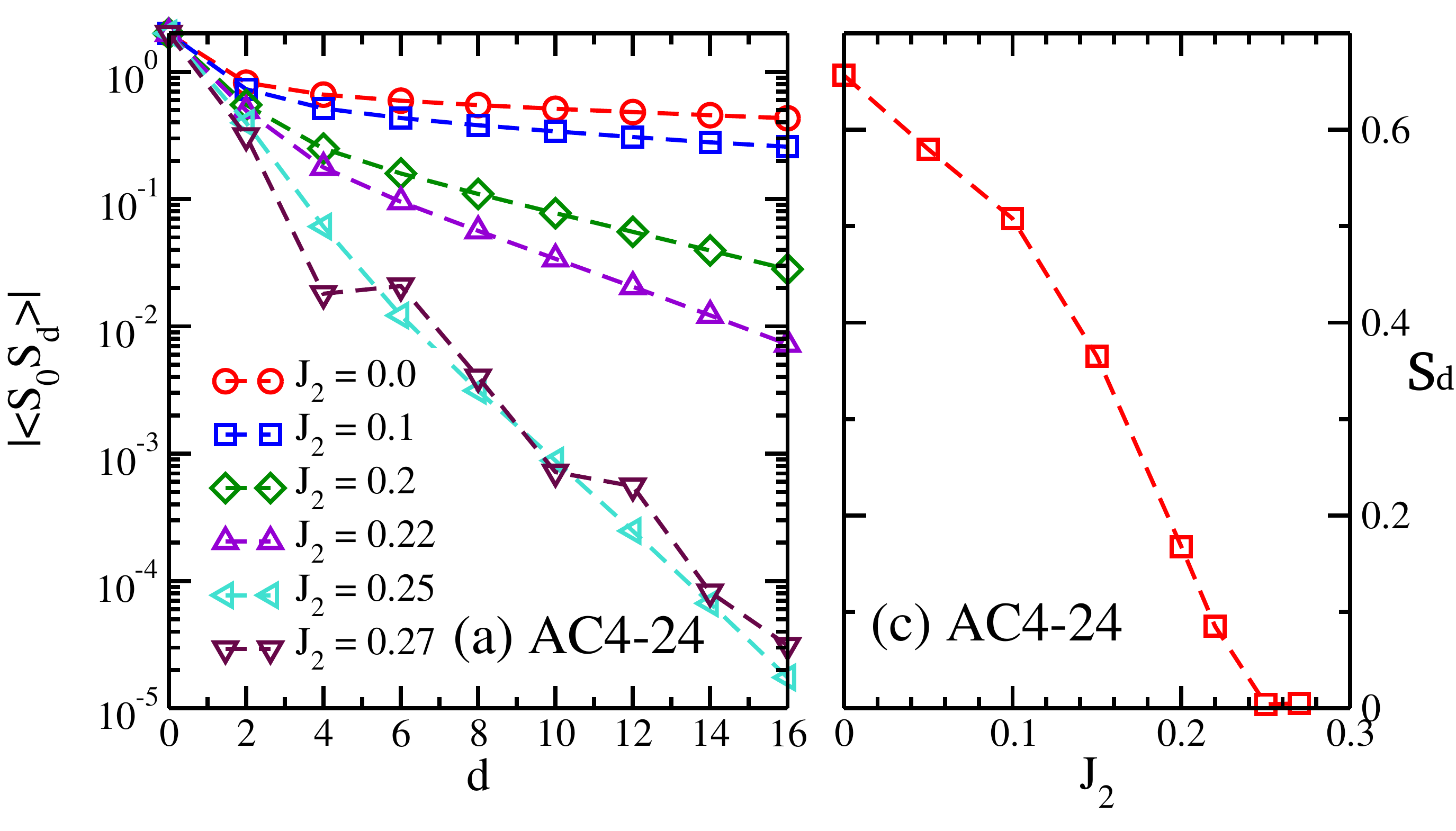}
  \includegraphics[width = 1.0\linewidth]{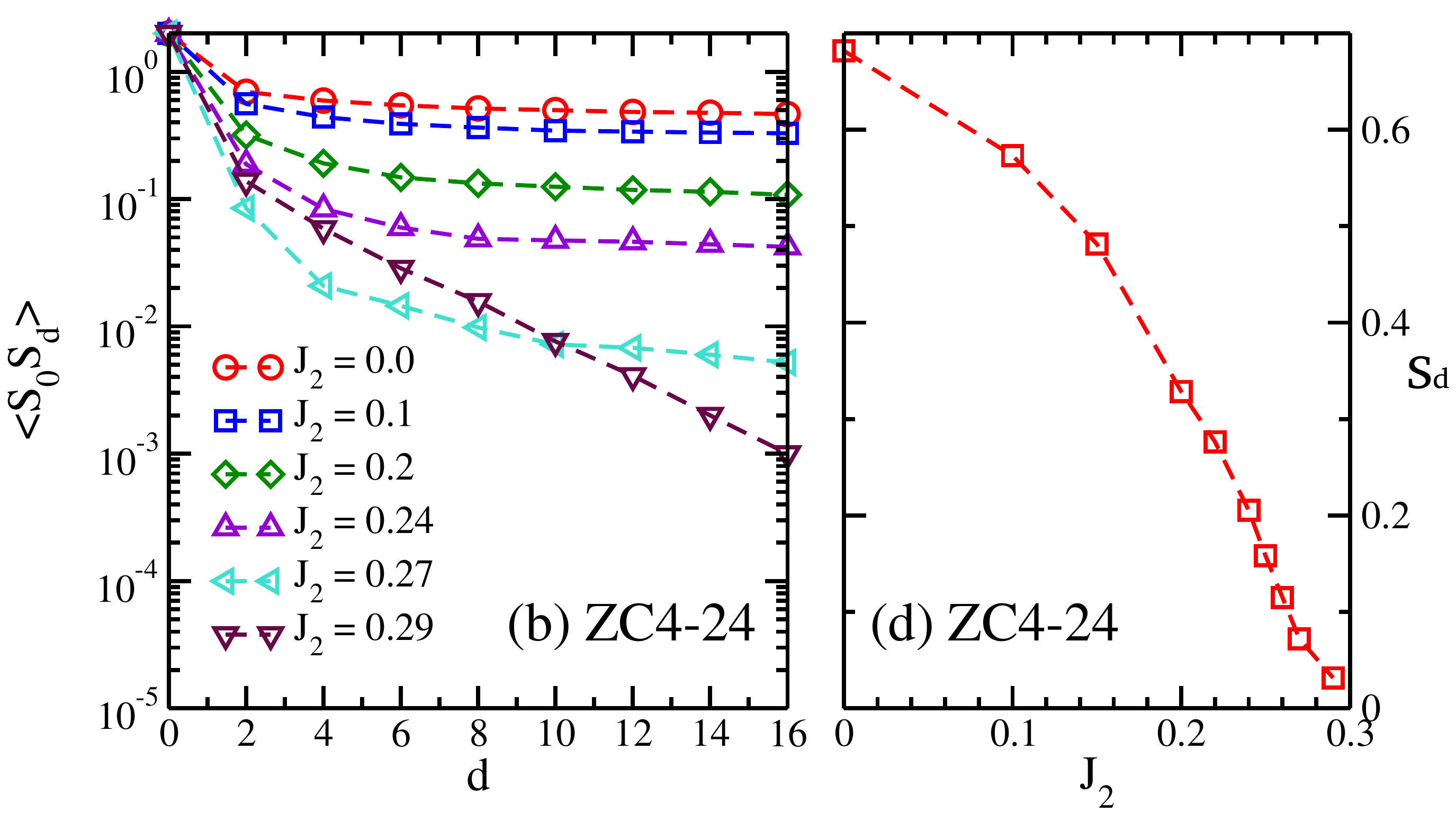}
  \caption{(a) and (b) are the log-linear plots of spin correlations on the AC4-24 and ZC4-24 cylinders.
	  (c) and (d) are the $J_2$ dependence of the long-distance spin correlations
	  $S_d \equiv \sqrt{|\langle S_0 S_d\rangle|}$ ($d$ is the longest distance in (a) and (b)).
	  In the subfigure (a), the correlation data for $J_2 < 0.27$ are all positive. For $J_2 = 0.27$,
	  the N\'{e}el AFM pattern is destructed, thus the correlations change sign in some places.} \label{spindecay_Neel}
\end{figure}

In Fig.~\ref{spinpattern_Neel}, we demonstrate the spin correlation functions
in real space for $J_2 = 0.0$ on the AC4-24 and ZC4-24 cylinders. Clearly, the spin
correlations exhibit a N\'{e}el AFM pattern with two magnetic sublattices.
We follow the $J_2$ dependence of the spin correlation decaying to detect 
the vanishing of N\'{e}el order. On the AC4-24 cylinder in Fig.~\ref{spindecay_Neel}(a), 
we find that the spin correlation length keeps decreasing with growing $J_2$,
which reaches a minimum at $J_2 \simeq 0.25$. Slightly above $J_2 = 0.25$ such as 
$J_2 = 0.27$ as shown in Fig.~\ref{spindecay_Neel}(a), the N\'{e}el AFM pattern 
of correlations is destructed, which signals a phase transition with vanishing the N\'{e}el order.
On the ZC4-24 cylinder as shown in Fig.~\ref{spindecay_Neel}(b), we find that
the spin correlations decay much slower than those on the AC4-24 cylinder
near the transition point, which indicates the finite-size effects of the system
when approching phase boundary. Beyond $J_2 \sim 0.28$, the spin correlations 
decay fast as shown in Fig.~\ref{spindecay_Neel}(b) at $J_2 = 0.29$. Based on
the spin correlations on both AC4 and ZC4 cylinders, we estimate the N\'{e}el
order vanishing at $J_2 \sim 0.26 - 0.28$.

Based on the decay behaviors of spin correlations, we plot the $J_2$
dependence of the long-distance spin correlations $S_d \equiv \sqrt{|\langle S_0 S_d\rangle|}$
($d$ is the longest distance in Figs.~\ref{spindecay_Neel}(a) and \ref{spindecay_Neel}(b))
as shown in Figs.~\ref{spindecay_Neel}(c) and \ref{spindecay_Neel}(d).
We find that the vanishing of $S_d$ and the destruction of the N\'{e}el AFM patter
on both geometries are consistent with a phase transition at $J_2 \simeq 0.26 - 0.28$.
We have also checked the spin correlations on AC6 and ZC6 cylinders. Although the long-range
correlations on these wider cylinders are not fully converged, their behaviors are qualitatively
consistent with those on the AC4 and ZC4 cylinders.

\section{Stripe AFM phase}

\begin{figure}
  \includegraphics[width=1.0\linewidth]{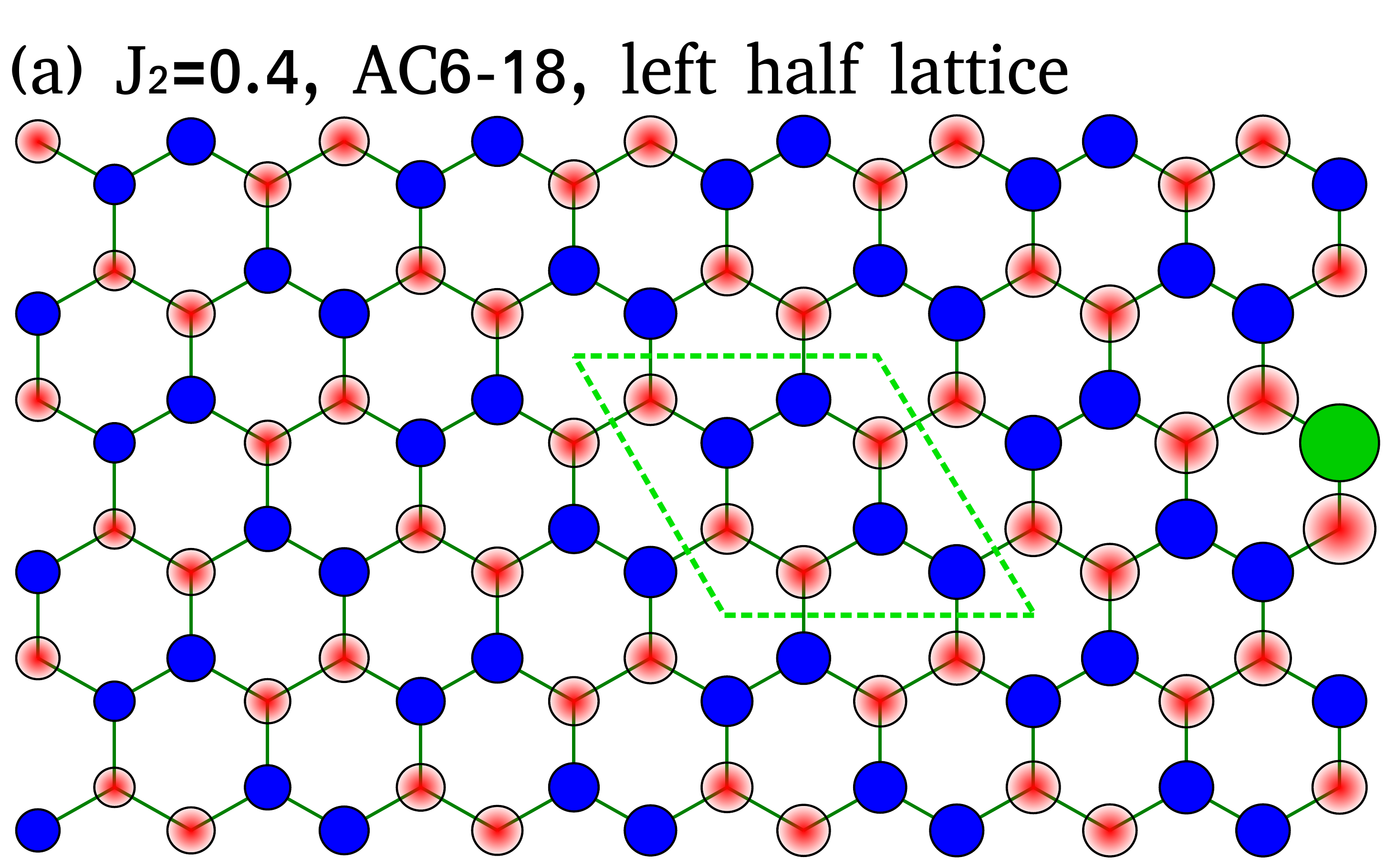}
  \includegraphics[width=1.0\linewidth]{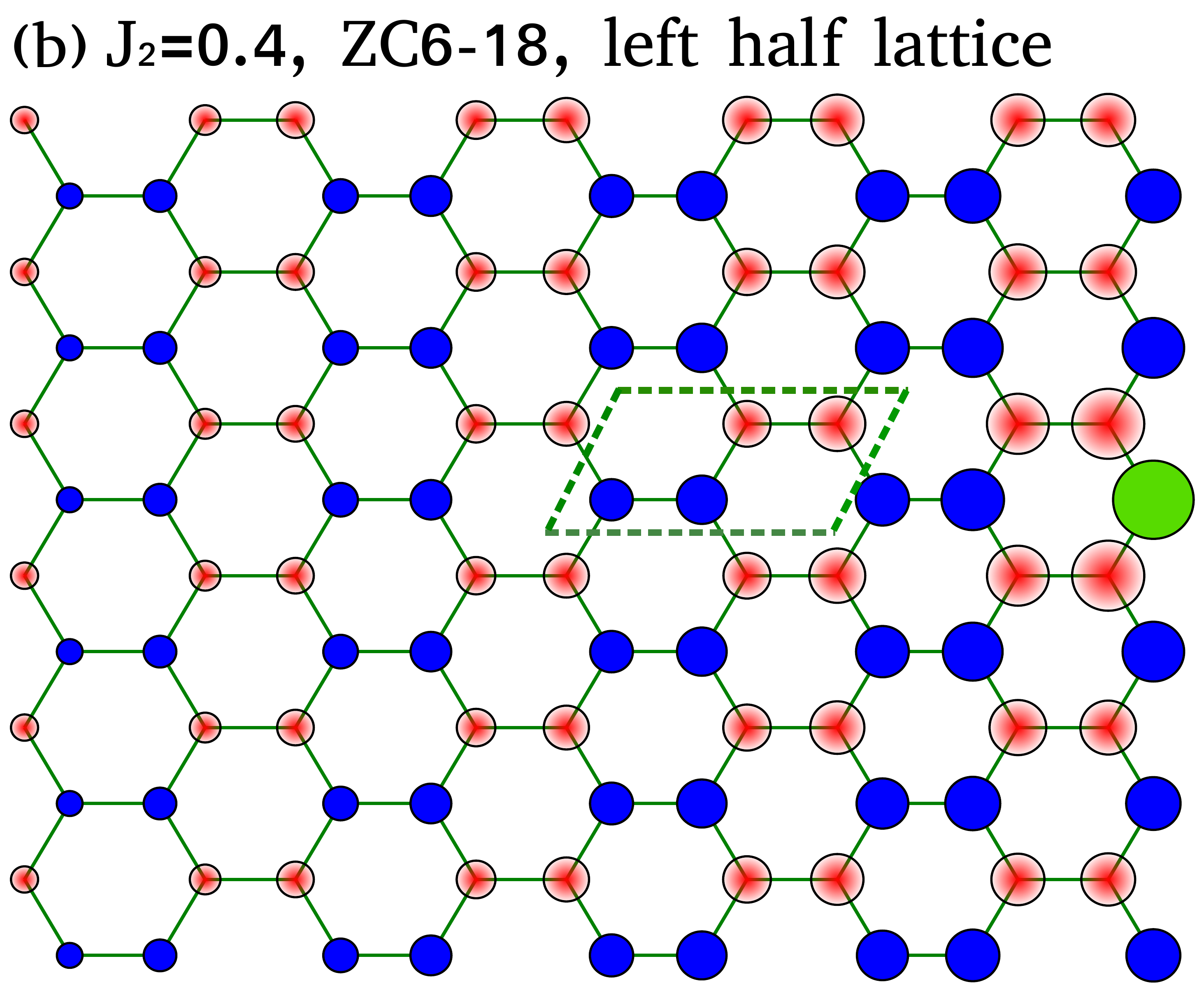}
  \caption{Spin correlation function $\langle \vec{S}_0 \cdot \vec{S}_j \rangle$ in real space for $J_2 = 0.4$
	   on (a) AC6-18 cylinder and (b) ZC6-18 cylinder. The green site is the reference spin $\vec{S}_0$ in the middle
 	   of lattice, and the blue solid and red shaded circle denote the positive and negative spin correlations, 
	   respectively. The area of the circle is proportional to the amplitude of correlations. The dashed 
	   rectangles denote the unit cells.} \label{stripe}
\end{figure}

\begin{figure}
  \includegraphics[width=1.0\linewidth]{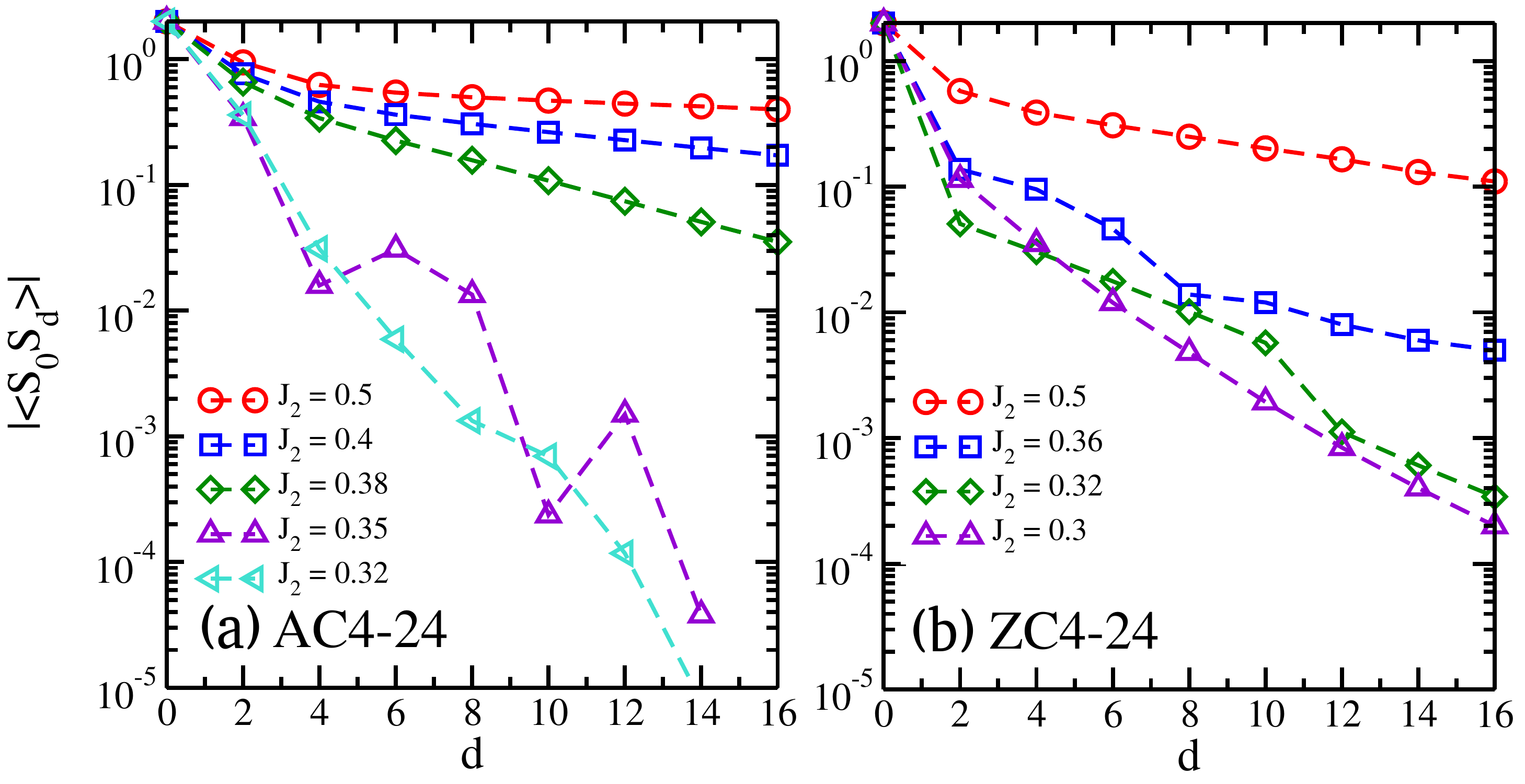}
  \includegraphics[width=1.0\linewidth]{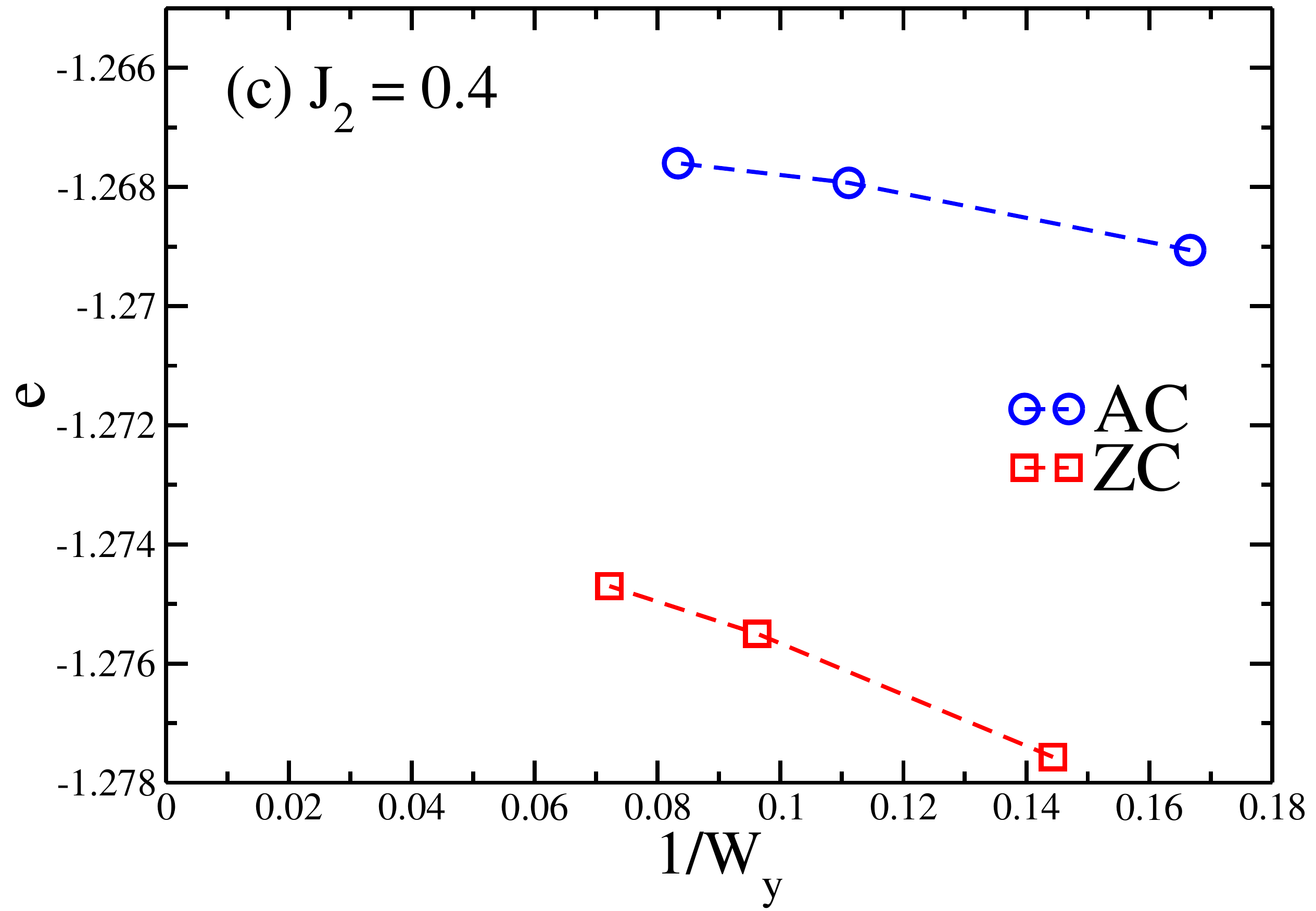}
  \caption{Log-linear plots of spin correlations at large $J_2$ side for (a) AC4-24 and (b)
	  ZC4-24 cylinders. (c) Cylinder width dependence of the bulk ground-state energy
	  for $J_2 = 0.4$ on the AC and ZC cylinders. ZC cylinders
	  always have the lower energy than AC cylinders.} \label{energy_04}
\end{figure}

In the spin-$1/2$ $J_1$-$J_2$ honeycomb Heisenberg model, the system is in a
staggered dimer phase with breaking lattice rotational symmetry and short-range
spin correlations for $J_2 \gtrsim 0.35$.
Interestingly, in this spin-$1$ system we find magnetically ordered states instead
of the staggered dimer.
As shown in Fig.~\ref{stripe}(a) for $J_2 = 0.4$ on the AC6-18 cylinder,
we find a magnetic ordered state with the $8$-site unit cell denoted by the
green dashed rectangle. This state is stable on all the 
AC cylinders that we have studied (AC4, AC6, and AC8).
However, on the ZC6-18 cylinder at $J_2 = 0.4$,
we find a stripe AFM state as shown in Fig.~\ref{stripe}(b). This stripe state
is also stable on the different ZC cylinders (ZC4, ZC6, and ZC8).
In Figs.~\ref{energy_04}(a) and \ref{energy_04}(b) of the log-linear plots of 
spin correlation functions at the large $J_2$ side, we find that 
the spin correlation length diminishes with decreasing $J_2$. On AC4 cylinder,
the spin correlations have a sharp increase for $J_2 \gtrsim 0.35$; and on ZC4 cylinder,
correlations grow rapidly for $J_2 \gtrsim 0.32$. For $J_2 \lesssim 0.32$, the spin correlations decay
quite fast to vanish, which is consistent with a non-magnetic phase region.

To identify which state is the exact ground state at large $J_2$ side,
we compare the ground-state energy on both cylinder geometries as shown in Fig.~\ref{energy_04}(c)
for $J_2 = 0.4$. We extract the ground-state energy from the bulk bond energy on long cylinder
systems. Interestingly, we find that the energies on ZC cylinder are always lower than
those on AC cylinder. As we expect, the two geometries should give the same energy
in the thermodynamic limit. Thus, the different energy in Fig.~\ref{energy_04}(c)
indicates the strong finite-size effects on systems we can study. Although we cannot
definitely determine which geometry gives the correct ground state in the thermodynamic limit,
based on our calculations we believe that the stripe state on ZC cylinder would win because it always has the lower
energy for  system sizes we studied.

\begin{figure}
  \includegraphics[width=1.0\linewidth]{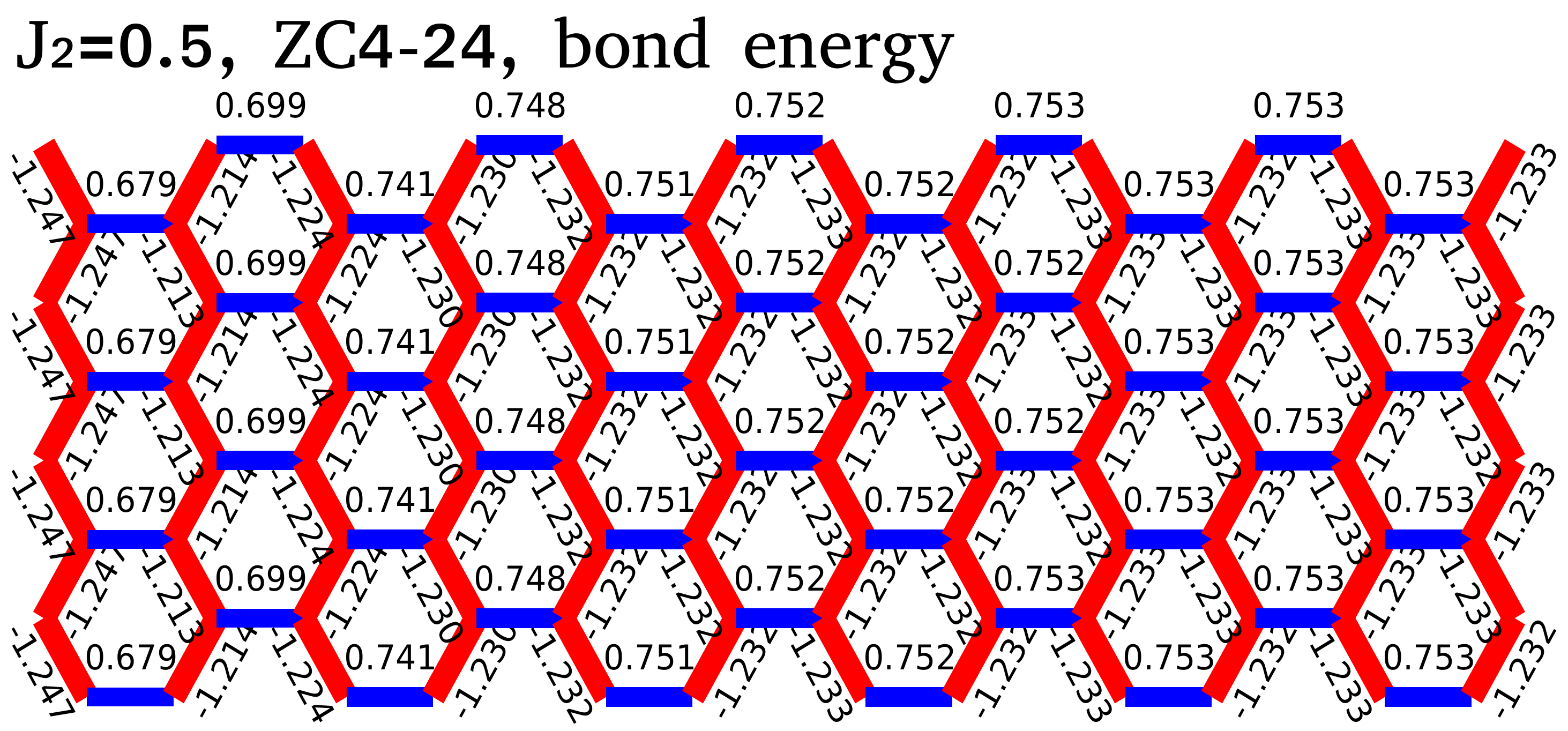}
  \caption{The nearest-neighbor bond energy for $J_2 = 0.5$ on the ZC4-24 cylinder. Here, we only show the left
	  half lattice. The positive horizontal (labeled by blue) and the negative vertical (labeled by red) bond energies
	  are consistent with the spin configuration in the stripe AFM state.} \label{bondenergy_05}
\end{figure}

As shown in Fig.~\ref{phasediagram} of the spin configuration in the stripe AFM phase, we
expect that the nearest-neighbor bond energy $\langle \vec{S}_i \cdot \vec{S}_j \rangle$ of the two parallel spins 
are positive and those between the anti-parallel spins are negative. This feature also can
be used to characterize the appearance of the stripe phase. In Fig.~\ref{bondenergy_05},
we demonstrate the bond energy for $J_2 = 0.5$ on ZC4-24 cylinder, which  indeed has
the bond energy pattern expected for the stripe AFM phase. With decreasing $J_2$,
we find that this bond energy pattern could persist to $J_2 = 0.32$, which is consistent with
the phase transition point estimated from spin correlation function.

\section{Intermediate phase region}

\begin{figure*}[t]
  \includegraphics[width=0.7\linewidth]{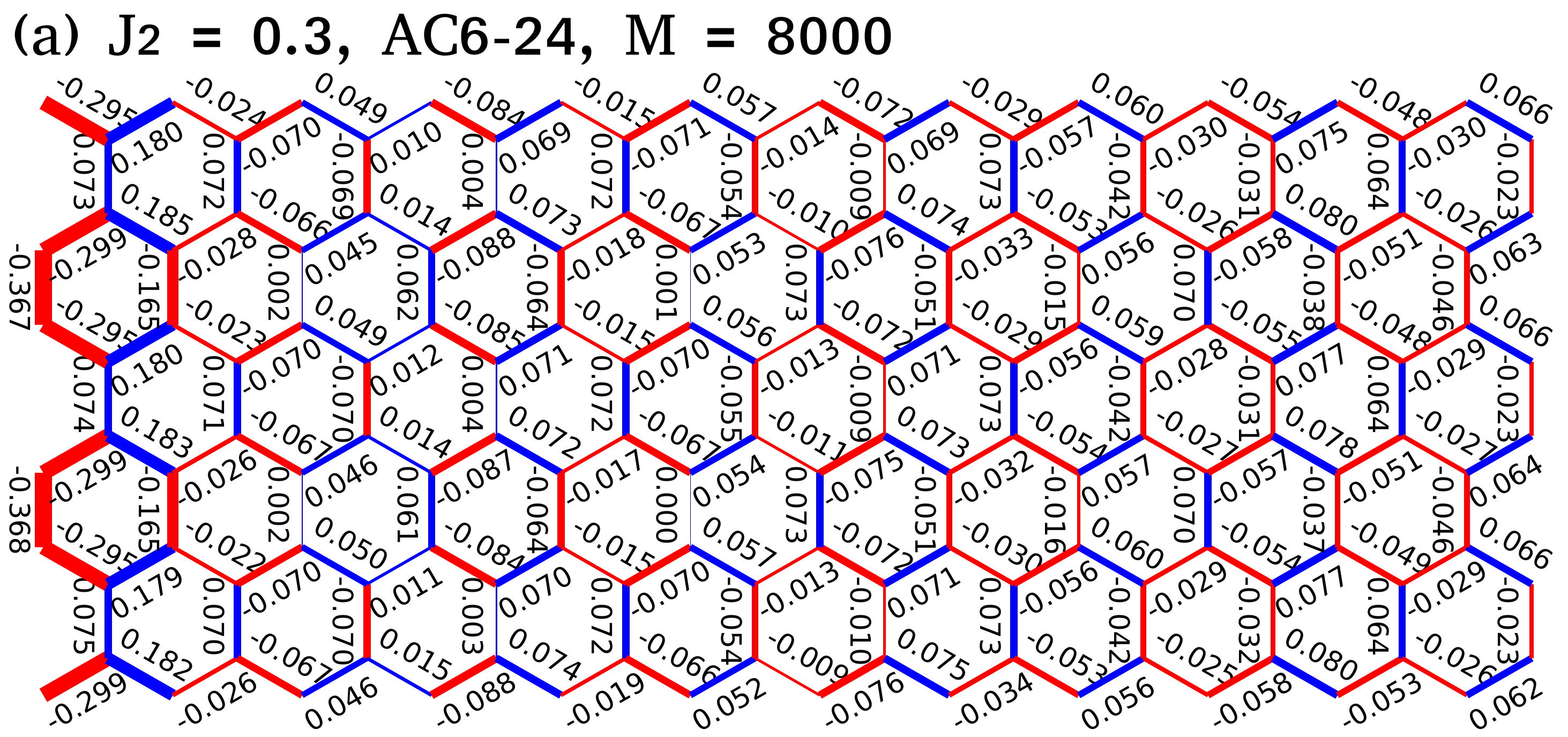}
  \includegraphics[width=0.7\linewidth]{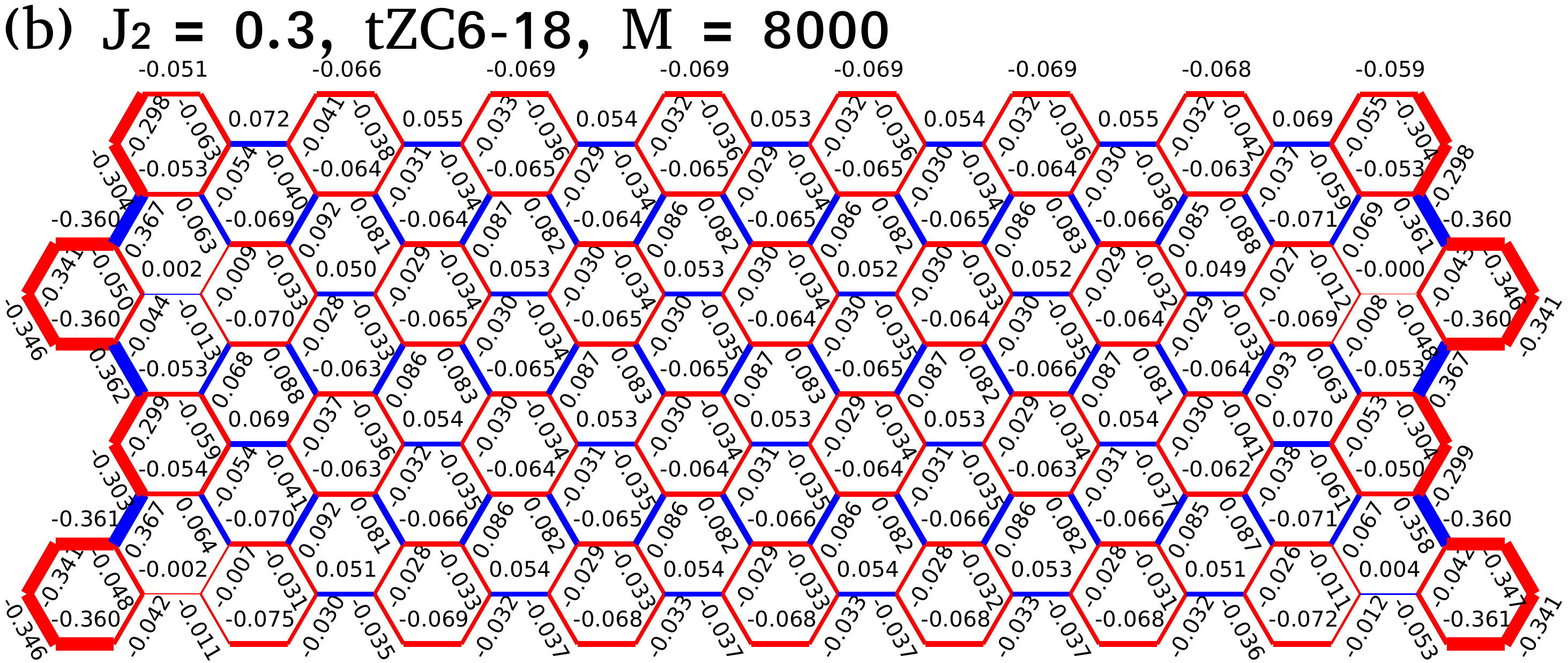}
  \caption{The nearest-neighbor bond energy textures for $J_2=0.3$ on (a) AC6-24 cylinder and (b) tZC6-18 cylinder, which are
	  obtained by keeping $8000$ $SU(2)$ states. The bond textures are obtained by subtracting the average
	  bond energy. The blue and red bonds denote the positive and negative bond textures, respectively. 
	  On both geometries, a PVB bond pattern is found. For AC6-24 in (a), only the left half
	  lattice is shown here.} \label{bondenergy}
\end{figure*}

Next, we investigate the intermediate region between the two magnetic ordered phases.
First of all, we calculate the first-neighbor bond energy $\langle S_i \cdot S_j \rangle$
to study the possible lattice symmetry breaking. To accommodate the possible valence-bond solid states
on cylinder system, we perform calculations on two cylinder geometries, the AC
cylinder and the trimmed ZC (tZC) cylinder with some trimmed sites on the open boundaries
(see the lattice in Fig.~\ref{bondenergy}(b)) \cite{PhysRevLett.110.127205,PhysRevB.88.165138}. 
On the AC4 and tZC4 cylinders, the bond energies are quite uniform in the bulk of cylinder, which is consistent with the
short order correlation length on narrow system. Thus, to detect the possible lattice symmetry breaking, 
we need go to the wider systems to study the behavior of correlation length with growing 
cylinder width \cite{PhysRevLett.110.127205, PhysRevB.88.165138, gong2014}.
As the DMRG convergence in the intermediate region is quite challenging on the wider systems,
we can only study the AC6 and tZC6 cylinders.
During the DMRG calculations, we measure the bond energy by increasing the optimal state number
step by step. By keeping the states up to $8000$ $SU(2)$ states, we find a PVB dimer pattern stabilized
as shown in Fig.~\ref{bondenergy}, which is robust in the whole lattice and thus has broken
the lattice translation symmetry on the finite-size cylinders. For the calculations on the AC6 and tZC6 cylinders by
keeping $8000$ states, the truncation errors are about $1\times 10^{-5}$.
Although we cannot obtain the more accurate results, based on the clearer PVB pattern with growing kept state number,
we argue that the PVB order is stable on these geometries. Interestingly, the field-theory analyses have
indicated that such a PVB state could emerge proximate to the N\'{e}el AFM phase on both the spin-$1/2$ and the 
spin-$1$ honeycomb Heisenberg models \cite{read1990}. We also study the bond energy for other $J_2$ around 
the intermediate region. For $J_2 = 0.31$, the systems also exhibit the PVB pattern. However, 
for $J_2 \geq 0.32$ the bond energy pattern appears like Fig.~\ref{bondenergy_05}, in consistent with the stripe state.

\begin{figure}
  \includegraphics[width=1.0\linewidth]{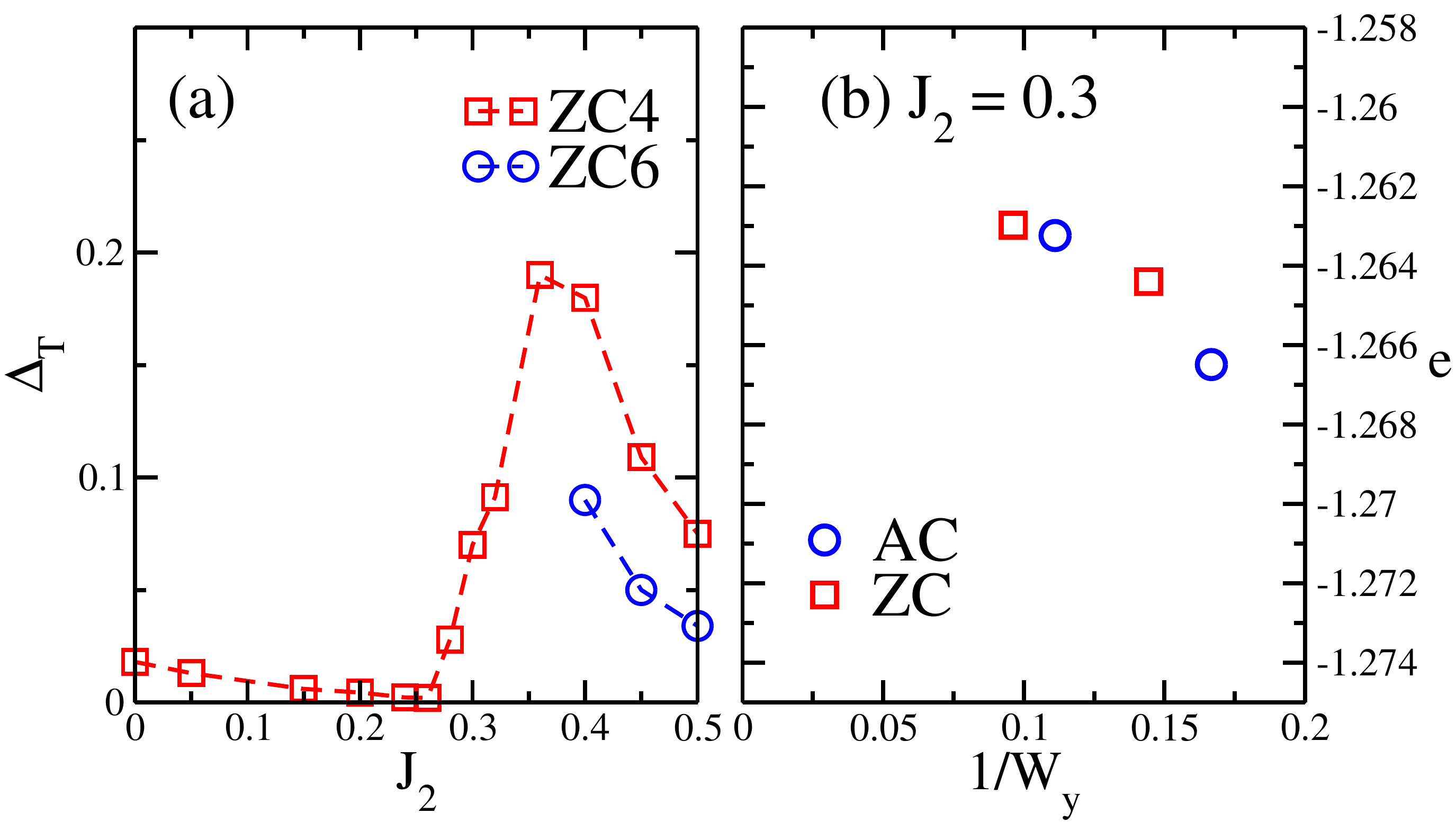}
  \caption{(a) $J_2$ dependence of the spin gap on the ZC4 and ZC6 cylinder systems.
	  (b) Cylinder width dependence of the bulk ground-state energy on AC (AC4 and AC6) and ZC (ZC4 and ZC6)
	  cylinders for $J_2 = 0.3$.} \label{gap}
\end{figure}

To further distinguish the three different phases, we calculate the spin gap
on long cylinder. As the broken $SU(2)$ symmetry and emerging Goldstone
boson, the spin gap is expected to vanish in both the N\'{e}el AFM and the stripe
AFM phases. However, as the spin singlet bonds formed in the valence-bond solid states,
the spin gap should be non-zero in a PVB state. To find the spin gap on cylinder,
we follow the standard method in DMRG calculations. We first sweep a long
cylinder to find the ground state with the energy $E_0$ , and then we sweep the
bulk by targeting the total spin $S = 1$ sector to find the lowest-energy state
with the energy $E_1$. Then the spin gap of the bulk $\Delta_{T}$ is the difference between
$E_0$ and $E_1$, $\Delta_T =  E_1 - E_0$. We calculate the long cylinder to make sure
the spin gap $\Delta_T$ is converged and independent of cylinder length.
We do not show the data on AC cylinders as their ground state in the large $J_2$
side has the higher energy on our studied systems. As shown in Fig.~\ref{gap}(a), 
the spin gap $\Delta_T$ is vanishing-small for $J_2 \lesssim 0.25$, which is consistent
with the N\'{e}el AFM order. At $J_2 \simeq 0.27$, the spin gap increases dramatically
and then decreases for $J_2 \gtrsim 0.35$. The enhanced spin gap in the intermediate region
could be served as an evidence for the valence-bond solid states.
In the stripe phase $J_2 \gtrsim 0.32$, the spin gap is large on ZC4 cylinder, 
which is related to  finite-size effects  because it should be vanished
in the thermodynamic limit. This size effect can be understood from the bond energy 
in Fig.~\ref{bondenergy_05}, which has the AFM bonds along the $y$ axis. Thus, the
spin triplet gap is strongly correlated  with the size scale in the $y$ direction,
which accounts for the large gap on the narrow ZC4 cylinder. To show the size dependence
of spin gap on the cylinder width, we also calculate the spin gap on the ZC6 cylinder 
deep inside the stripe phase. As expected, the spin gap on ZC6 cylinder drops rapidly, consistent with
the vanished gap in the thermodynamic limit. We do not show the spin gap on ZC6 cylinder
for the intermediate region because the DMRG calculations in the spin-$1$ sector 
are quite far from convergence in this region.

Finally, we demonstrate the bulk ground-state energy for $J_2 = 0.3$ on different cylinders.
We obtain the bulk energy from calculating all the bond energy in the middle of cylinder. 
As shown in Fig.~\ref{gap}(b), the ground-state energy at $J_2 = 0.3$ smoothly increases with growing cylinder width, which
behaves differently from the energy scaling in the stripe phase demonstrated in Fig.~\ref{energy_04}(c),
where the energies on the two geometries scale separately. We also notice that the ground-state
energy per site at $J_2 = 0.3$ changes slightly with growing system width.
The energy appears to approach $e_{\infty} \simeq -1.262$, which provides an
upper bound for the ground state energy in the thermodynamic limit.

\section{Entanglement entropy and spectrum}

\begin{figure}
  \includegraphics[width=0.8\linewidth]{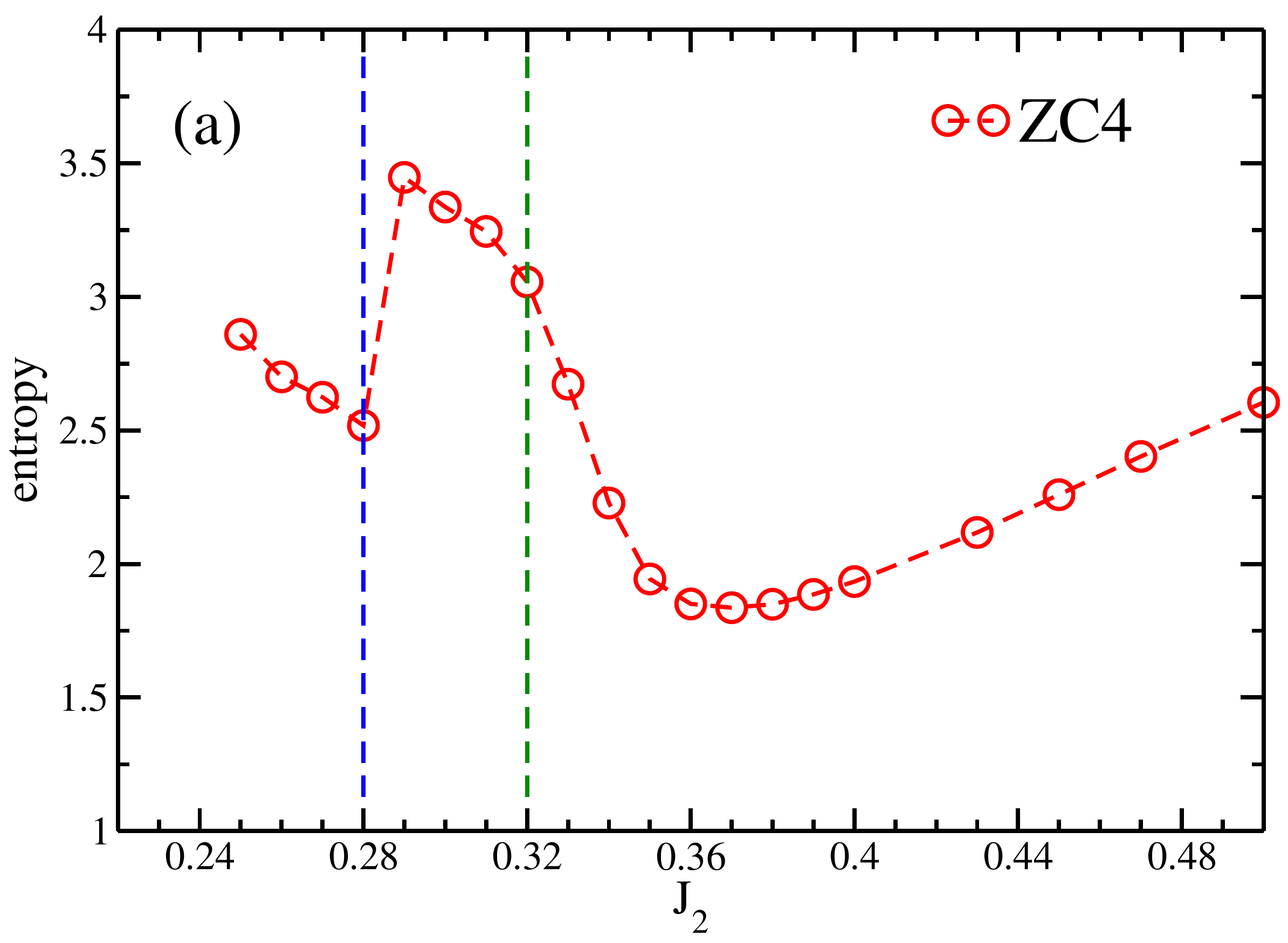}
  \includegraphics[width=0.98\linewidth]{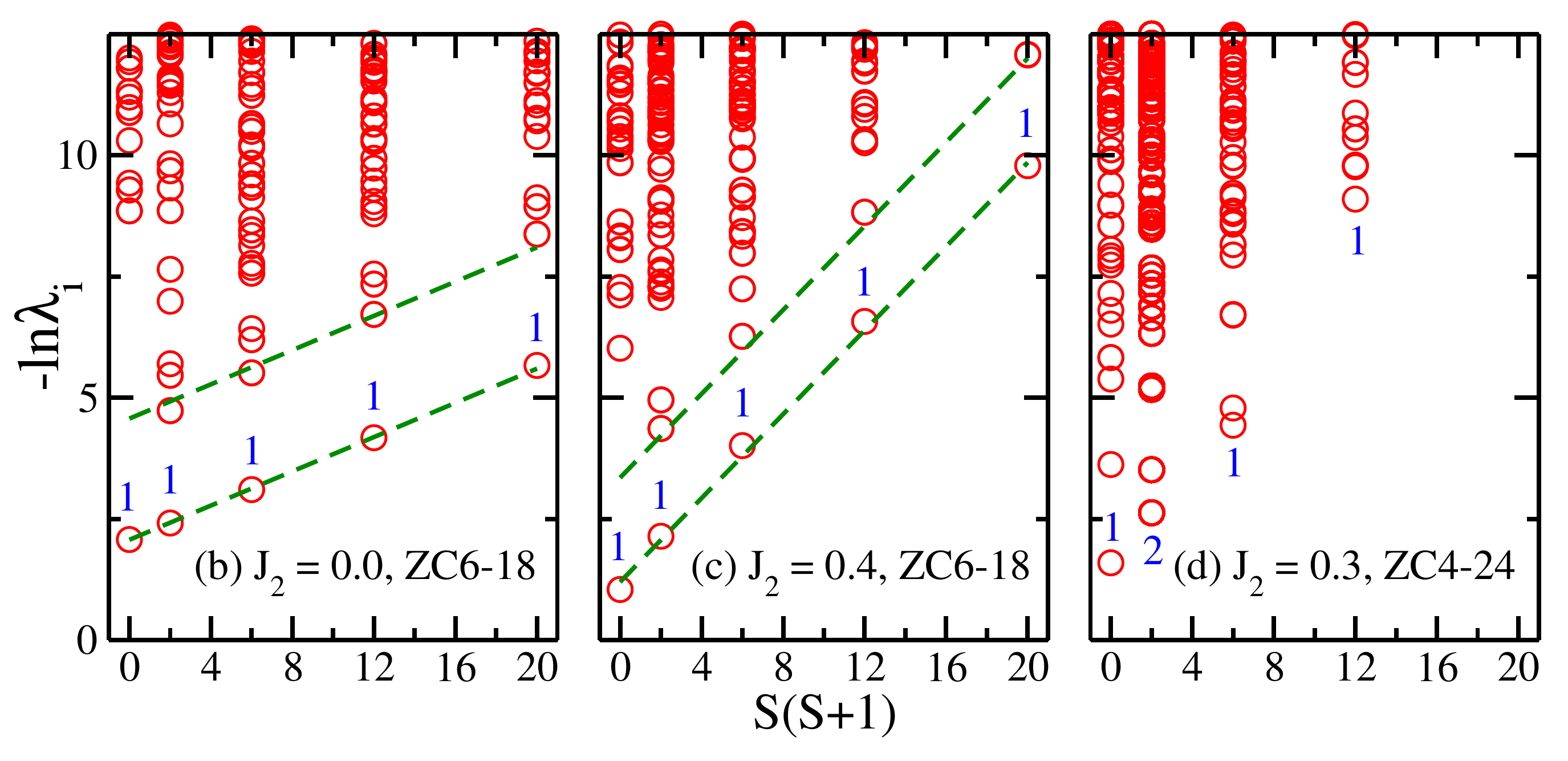}
  \caption{(a) $J_2$ dependence of the bipartite entanglement entropy on the ZC4 cylinder. At $J_2 \simeq 0.28$ and $0.32$, 
	the entropy has the sharp jump and drop, which appear consistent with the quantum phase transitions. 
	Entanglement spectra for (b) $J_2 = 0.0$ on ZC6-18 cylinder, (c) $J_2 = 0.4$ on ZC6-18 cylinder, and (d) $J_2 = 0.3$ 
	on ZC4-24 cylinder. The blue numbers denote the number of the largest eigenvalues in each $S$ sector. The green dashed
	lines denote the tower of states structure in the magnetic ordered states.} \label{entropy}
\end{figure}

To further characterize the different phases and phase transitions, we study the bipartite 
entanglement entropy and entanglement spectrum. In Fig.~\ref{entropy}(a), we demonstrate
the $J_2$ dependence of entropy on the ZC4 cylinder. The entropy shows a sharp jump at $J_2 \simeq 0.28$ 
and a drop at $J_2 \simeq 0.32$, which have been shown to characterize various phase transitions 
in one-dimensional systems \cite{chen2006, deng2006, kargarian2008, liu2012}
and are consistent with the identified transitions points.
While the discontinuous jump at $J_2 \simeq 0.28$ suggests a first-order transition from the 
N\'{e}el AFM to the non-magnetic phase, the smoother entropy decrease near $J_2 \simeq 0.32$ 
might be consistent with a weak first-order transition.

In Figs.~\ref{entropy}(b-d), we demonstrate the bipartite entanglement spectrum in each phase.
For the ordered phases with continuous symmetry breaking, the lower part of the entanglement spectrum 
is in correspondence with the ``tower of states" (TOS) spectrum \cite{metlitski2011, alba2013, kolley2013}.
In Figs.~\ref{entropy}(b) and \ref{entropy}(c) of the spectra in both the magnetic ordered states with
breaking $SU(2)$ to $U(1)$ symmetry, the spectra have a single dominant eigenvalue in each $S$ sector,
which are separated from the higher levels by the entanglement gap and follow a linear behavior with $S(S+1)$
($S$ is the quantum number of total spin). All these features are consistent with the TOS structures of 
the energy spectra in the corresponding magnetic ordered states on the honeycomb lattice \cite{fouet2001}. 
In the intermediate phase region, 
the spectrum is totally different from the magnetic ordered states, as shown in Fig.~\ref{entropy}(d) for
$J_2 = 0.3$ on ZC4 cylinder \cite{note}.
The low-lying degeneracy in the $S=1$ sector changes from $1$ to $2$, and
there is no clear  entanglement gap between the largest eigenvalues and the rest of spectrum.
These features distinguish the intermediate phase from the magnetic ordered phases.

\section{Summary}

We have studied the quantum phase diagram of the spin-$1$ $J_1$-$J_2$
Heisenberg model on the honeycomb lattice using
density-matrix renormalization group calculations on cylinder system.
We have established three different phases including two magnetic ordered 
phases and a non-magnetic phase. For $J_2 \lesssim 0.27$,  we find a N\'{e}el AFM phase.
For $J_2 \gtrsim 0.32$, we find two possible candidate magnetic ordered states
depending on the different geometries.
On AC cylinders, the system is a magnetic order state with the $8$-site unit cell,
while on ZC cylinders it is a stripe AFM state. By comparing the bulk ground-state
energy on the two geometries, we find that the ZC cylinders always have the lower
energy than the AC cylinders, which strongly suggests the stripe AFM state as
the true ground state in the thermodynamic limit. 

Between these two magnetically ordered phases with $0.27 \lesssim J_2 \lesssim 0.32$, we find a non-magnetic phase
region. On both AC6 and tZC6 cylinders, the systems have the non-uniform bond energy.
By increasing the kept states to $8000$ $SU(2)$ states (equivalent to about $24000$ $U(1)$ states), 
we find a stable plaquette valence-bond order emerging in the systems.
The spin gap on finite-size cylinder also enhances dramatically in this phase region. 
Our results indicate that the plaquette state is a strong candidate for this non-magnetic phase.
Moreover, the sizable entropy change on the phase boundaries indicates that
the nature of the phase transitions from the magnetic ordered phases to the non-magnetic phase
might be first  order.
\\

\section*{ACKNOWLEDGEMENTS}
We thank T.~Senthil and F.~Wang for stimulating discussions.
This research is supported by the National Science Foundation through 
grants DMR-1205734 (S.S.G.), DMR-1408560 (D.N.S.), and the U.S.
Department of Energy, Office of Basic Energy Sciences under grants No. DE-FG02-06ER46305 (W.Z.).

\bibliographystyle{apsrev}
\bibliography{spin_1_honeycomb}{}

\begin{thebibliography}{80}
\bibitem[{not()}]{note}
\bibinfo{note}{We do not show the entanglement spectrum for ZC6 cylinder
  because compared with bond energy, it is harder to obtain the converged
  entanglement spectrum on ZC6 cylinder.}
\end{thebibliography}

\end{document}